# The "Vector-Model" Wavefunction: spatial description and wavepacket formation of quantum-mechanical angular momenta

T. Peter Rakitzis[*], Michail E. Koutrakis, and George E. Katsoprinakis

*Department of Physics, University of Crete, 70013 Heraklion-Crete, Greece*

*IESL-FORTH, N. Plastira 100, Heraklion-Crete 71110, Greece*

[*]*Email:* ptr@iesl.forth.gr

In quantum mechanics, spatial wavefunctions describe distributions of a particle's position or momentum, however, wavefunctions are not used to describe the spatial distribution of the angular momentum vector $\boldsymbol{j}$. Here, we introduce the concept of an asymptotic spatial angular-momentum wavefunction, $X_m^j(\varphi, \theta, \chi) = e^{im\varphi} \, \delta(\theta - \theta_m) \, e^{i\left(j+\frac{1}{2}\right)\chi}$, which treats $\boldsymbol{j}$ in the $|jm\rangle$ state as a three-dimensional entity; $\varphi$, $\theta$, $\chi$ are the Euler angles, and $\theta_m$ is the Vector-Model polar angle, given by $\cos\theta_m = m/|\boldsymbol{j}|$. A wealth of geometric information about $\boldsymbol{j}$ can be deduced from the eigenvalues and spatial transformation properties of the $X_m^j(\varphi, \theta, \chi)$. Specifically, the $X_m^j(\varphi, \theta, \chi)$ wavefunctions give a computationally simple description of the sizes of the particle and orbital-angular-momentum wavepackets (constructed from Gaussian distributions in $j$ and $m$), given by effective wavepacket angular uncertainty relations for $\Delta m \Delta \varphi$, $\Delta j \Delta \chi$, and $\Delta \varphi \Delta \theta$. The $X_m^j(\varphi, \theta, \chi)$ also predict the position of the particle-wavepacket angular motion in the orbital plane, so that the particle-wavepacket rotation can be experimentally probed through continuous and non-destructive $\boldsymbol{j}$-rotation measurements. Finally, we use the $X_m^j(\varphi, \theta, \chi)$ to determine geometrically well-known asymptotic expressions for Clebsch-Gordan coefficients, Wigner $d$-functions, the gyromagnetic ratio of elementary particles, $g = 2$, and the $m$-state-correlation matrix elements. Interestingly, for low $j$, even down to $j = 1/2$, these expressions are either exact (the last two) or excellent approximations (the first two), showing that the $X_m^j(\varphi, \theta, \chi)$ give a useful spatial description of quantum-mechanical angular momentum, and provide a smooth connection with classical angular momentum.

## I. INTRODUCTION

In quantum mechanics, the spatial wavefunction of a particle in a central field is described by $\Psi(r,\theta,\varphi) = \psi(r)Y_M^L(\theta,\varphi)$, where $Y_M^L(\theta,\varphi)$ are spherical harmonics, $L$ is the angular momentum quantum number of the angular momentum $\boldsymbol{L}$, and $M$ is the projection of $\boldsymbol{L}$ along the $Z$ axis, which ranges from $-L$ to $L$ in integer steps [1, 2]. The constraint on the range of allowed values of $M$ stems from the requirement that the wavefunction be normalizable [2]. Subsequently, the angular position of the particle is described by $|Y_M^L(\theta,\varphi)|^2$, and expectation values of operators can be calculated.

In contrast, the orbital or spin angular momentum of a particle is not usually associated with a spatial wavefunction, so that the angular distribution of the angular momentum $\boldsymbol{j}$ cannot be calculated in an analogous way to that of the position of the particle. The lack of a spatial wavefunction for angular momentum $\boldsymbol{j}$ makes less clear both its visualization and the transition to classical mechanics.

Here, we identify an asymptotic ($j \to \infty$) spatial wavefunction $X_m^j(\varphi,\theta,\chi)$ of the $|jm\rangle$ state, which is the "missing link" between quantum and classical angular momentum:

$$X_m^j(\varphi,\theta,\chi) = e^{im\varphi}\,\delta(\theta-\theta_m)\,e^{i\left(j+\frac{1}{2}\right)\chi} \tag{1}$$

where $\varphi$, $\theta$, $\chi$ are the Euler angles, with $\varphi$ and $\chi$ being azimuthal angles about $Z$ and $z'$, respectively, $\theta$ is the polar angle, and $\theta_m$ is the Vector-Model (VM) polar angle given by $cos\theta_m = m/|\boldsymbol{j}|$. Here, $Z$ is the space-fixed quantization axis, and $z'$ is the body-fixed axis, which coincides with the direction of $\boldsymbol{j}$. The wavefunction $X_m^j(\varphi,\theta,\chi)$, depicted in Fig. 1, is an extension of the traditional VM (Fig. 1), with all the advantages, [1,2,3,4,5,6,7], but without its limitations [8]. It can also be viewed as an asymptotic eigenfunction of the space-fixed frame, $\hat{J}^2, \hat{J}_Z$, and body-fixed frame, $\hat{J}_{z'}$, angular-momentum operators, and satisfies the differential equations involving $\hat{J}_\pm$ and $\hat{J}_\mp$ [9,10], in the reduced VM-space space for fixed $\theta = \theta_m$, so that it is in the form of an asymptotic symmetric-top wavefunction.

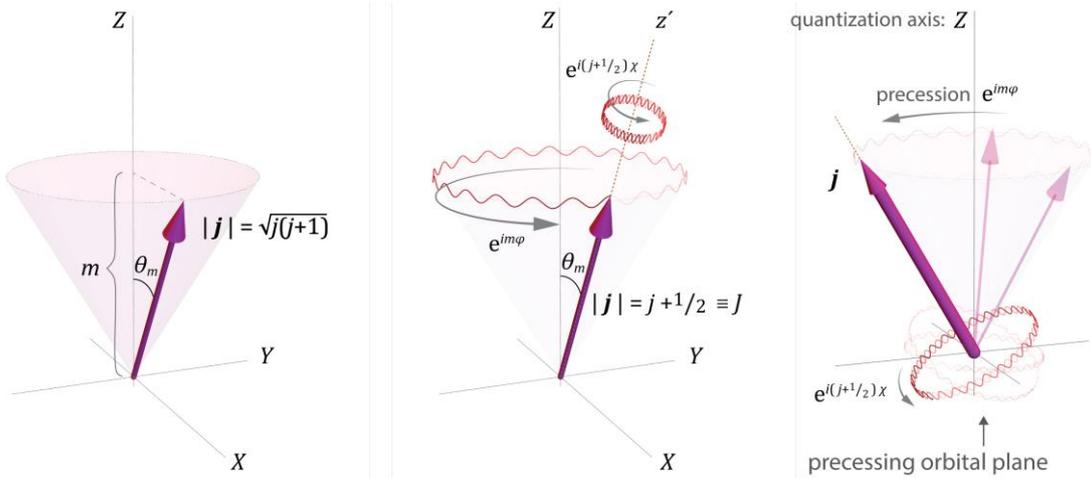

*Fig. 1 (a) The Vector Model representation of state $|jm\rangle$: vector $\boldsymbol{j}$, with modulus $|\boldsymbol{j}| = \sqrt{j(j+1)}$, precesses about the Z axis at angle $\theta_m$, with constant projection $m$. (b) Depiction of the VM wavefunction, showing a 2-dimensional standing wave, where we plot $Re[X_m^j(\varphi,\theta,\chi)]$ for integer $m$, and the transformation under rotation of $\varphi$ about Z, and $\chi$ about $\boldsymbol{j}$. (c) A charged particle orbits perpendicular to $\boldsymbol{j}$, and the orbital plane precesses about the quantization axis Z.*

The most unusual feature of wavefunction $X_m^j(\varphi, \theta, \chi)$ is that the projection of the angular momentum $\boldsymbol{j}$ along the $z'$ axis is $\left(j + \frac{1}{2}\right)$. This makes sense physically from the perspective that it is the asymptotic $(j \to \infty)$ length $|\boldsymbol{j}| = \sqrt{j(j+1)} \to j + \frac{1}{2}$; however, this projection is different from the usual allowable values (even beyond the usual range of allowed values), and therefore it violates the condition needed for normalizable wavefunctions. The consequence is that $X_m^j(\varphi, \theta, \chi)$ should not be normalizable (which is the case, as it is proportional to a delta function) and that expectation values of operators cannot be calculated. Nonetheless, we present the case that this wavefunction can give important information on the spatial properties of $\boldsymbol{j}$ in the $|jm\rangle$ state, and that this information is complementary to the angular information obtained from the spatial wavefunction $Y_M^L(\theta, \varphi)$ of the associated particle.

We stress that the $X_m^j(\varphi, \theta, \chi)$ is an auxiliary wavefunction, in that it adds an additional geometrical description to the $|jm\rangle$ state. For orbital angular momentum, the $Y_M^L(\varphi, \theta)$ describe the angular distribution of the particle, whereas the $X_M^L(\varphi, \theta, \chi)$ describe the angular distribution of the associated angular momentum vector, $\boldsymbol{L}$; such a geometrical description of $\boldsymbol{L}$ is absent in standard quantum mechanics. Asymptotically, the angular distribution of $\boldsymbol{L}$ has clear physical meaning, and wavepackets of $X_M^L(\varphi, \theta, \chi)$ correctly describe the transition to classical mechanics, with $\boldsymbol{L}$ becoming a well-defined classical vector; interestingly, the $X_m^j(\varphi, \theta, \chi)$ give geometrical predictions that are useful down to low $j$. For spin $S$, a counterpart to the $Y_M^L(\varphi, \theta)$ does not exist, however the $X_m^S(\varphi, \theta, \chi)$ can still be used to derive geometrical information about spin, as we discuss in later Sections.

The $X_m^j(\varphi, \theta, \chi)$ "Vector-Model" wavefunction (VMW) allows us to treat the $|jm\rangle$ state as a three-dimensional entity: a semiclassical vector $\boldsymbol{j}$ at angle $\theta_m$ to the $Z$ axis with projection $m$, and with rotational transformations about $Z$ and $z'$ given by $e^{im\varphi}$ and $e^{i(j+1/2)\chi}$, respectively. Although the $X_m^j(\varphi, \theta, \chi)$ is exact only in the ideal high-$j$ limit, we show that the concept of a spatial wavefunction of $\boldsymbol{j}$ is useful for low $j$ as well. Most importantly, the $X_m^j(\varphi, \theta, \chi)$ now allows the quantitative prediction of the properties of the coupled angular-momentum and particle wavepackets, with a straightforward physical picture of the emergence and orientation of the orbital plane, and of the angular size of the wavepackets, given by uncertainty relations for $\Delta m \Delta \varphi$, $\Delta j \Delta \chi$, and $\Delta \varphi \Delta \theta$. In contrast, exact quantum-mechanical $j$-wavepackets lack a clear physical picture with predictive power, and are computationally intensive, especially for high $j$ [scaling as the $(2j+1)^2$ density-matrix elements]. In addition, the three-dimensional $X_m^j(\varphi, \theta, \chi)$ wavepackets rotate to mirror the particle wavepacket rotation, such as for Kepler orbits [11]. This rotation, not observable in classical mechanics or for single quantum states, allows nondestructive experimental probing of particle-wavepacket rotation through $j$-rotation measurements.

The $X_m^j(\varphi, \theta, \chi)$ is not to be confused with the coherent spin state representation [12,13], with which it shares some similarities (such as a similar dependence on the angles $\theta$ and $\varphi$), but also has some crucial differences. The main novel feature of the $X_m^j(\varphi, \theta, \chi)$ is the existence of the body-fixed projection $\left(j + \frac{1}{2}\right)$, i.e., the dependence on the angle $\chi$, making it a symmetric-top solution of the angular momentum operators. The semiclassical interpretation of the dependence of the $X_m^j(\varphi, \theta, \chi)$ on the angles $\varphi, \theta, \chi$ is clear, and we describe it with respect to an orbiting charged particle in Fig. 1c: the term $\delta(\theta - \theta_m)$ describes the orientation of the orbital plane (and $\boldsymbol{j}$) to the Z axis; the $e^{i\left(j+\frac{1}{2}\right)\chi}$ term describes the phase of the particle in the orbital plane, as a function of the angle $\chi$ about the $z'$ axis (parallel to $\boldsymbol{j}$), which produces a magnetic moment along $\boldsymbol{j}$; and the $e^{im\varphi}$ term describes the phase of the particle due to precession of the orbital plane (and $\boldsymbol{j}$) about the quantization axis Z by angle $\varphi$. Therefore, it is clear that each of the three terms above play indispensable roles in the connection of quantum and classical mechanics, as they each correspond to clear classical motions or geometries.

The $X_m^j(\varphi, \theta, \chi)$ then give clear geometrical interpretations, and exact values or excellent approximations, for the important quantities in quantum-mechanical angular momentum: Clebsch-Gordan (C-G) coefficients, Wigner $d$-functions, the gyromagnetic ratio of elementary particles $g = 2$, and the $m$-state correlation matrix elements $\langle j_3 m_3 \mid j_{1X} j_{2X} \mid j_3 m_3 \rangle$. We conclude that quantum-mechanical angular momentum is well approximated by a clear semiclassical picture, even for low $j$, which is a surprising result, in contrast to how quantum-mechanical angular momentum is usually taught or understood. In fact, the success of this asymptotic geometrical wavefunction has prompted work to propose exact low-$j$ wavefunctions, and to investigate their usefulness [14].

Finally, spin networks are important to several fields [15], including atomic and molecular physics [10], quantum computation [16,17], and quantum gravity [18,19,20]. The determination and interpretation of asymptotic expressions for Wigner $3nj$ symbols of high $n$ in spin networks, which arise in the coupling of $n + 1$ angular momenta [21], will be aided by the use of the $X_m^j(\varphi, \theta, \chi)$.

## II. VM WAVEFUNCTION

The $X_m^j(\varphi, \theta, \chi)$ wavefunction is in the form of an asymptotic $(j \to \infty)$ symmetric-top wavefunction or Wigner rotation matrix element [10], so that $X_m^j(\varphi, \theta, \chi) \equiv |j \, J \, m\rangle \equiv D_{m \, J}^{j \, *}(\varphi, \theta, \chi)$, where $J = |\boldsymbol{j}| = j + 1/2$. The main difference with a usual symmetric-top $|j \, m' \, m\rangle$ state is that the body-fixed projection $m' = j + \frac{1}{2}$ is equal to the magnitude of $|\boldsymbol{j}|$, so that $\boldsymbol{j}$ points exactly along the body-fixed axis $z'$, projecting fully along $z'$. In the limit of quantized $j \to \infty$, where $|\boldsymbol{j}| = \sqrt{j(j+1)} \to j + \frac{1}{2}$ is an integer or half integer, the $e^{i\left(j+\frac{1}{2}\right)\chi}$ term gives a standing wave about $z'$. Note that, $X_m^j(\varphi, \theta, \chi)$ is proportional to a Dirac delta-function and is thus not normalizable, but geometric information will be inferred from the eigenvalues and spatial transformation properties of $X_m^j(\varphi, \theta, \chi)$.

We demonstrate that $X_m^j(\varphi, \theta, \chi) \equiv |j \, J \, m\rangle$ is an asymptotic eigenvector of the operators, $\hat{\boldsymbol{J}}^2, \hat{J}_Z$, and $\hat{\boldsymbol{j}}^2, \hat{j}_{z'}$, as well as an asymptotic solution to differential equations involving $\hat{J}_\pm$ and $\hat{j}_\mp$, for the VM space of reduced dimensionality given by $\theta = \theta_m$ (the operators and calculations are given in Appendix A), and we compare to expressions for the standard symmetric-top wavefunctions $|j \, m' \, m\rangle$ [10]:

$$\hat{J}^2|j\,m'\,m\rangle = \hat{j}^2|j\,m'\,m\rangle = j(j+1)|j\,m'\,m\rangle \qquad \rightarrow \hat{J}^2|j\,J\,m\rangle = \hat{j}^2|j\,J\,m\rangle = J^2|j\,J\,m\rangle \qquad (2a)$$

$$\hat{J}_Z|j\,m'\,m\rangle = m|j\,m'\,m\rangle \qquad\qquad\qquad \rightarrow \hat{J}_Z|j\,J\,m\rangle = m|j\,J\,m\rangle \qquad (2b)$$

$$\hat{j}_{z'}|j\,m'\,m\rangle = m'|j\,m'\,m\rangle \qquad\qquad\qquad \rightarrow \hat{j}_{z'}|j\,J\,m\rangle = J|j\,J\,m\rangle \qquad (2c)$$

$$\hat{J}_\pm|j\,m'\,m\rangle = \sqrt{j(j+1)-m(m\pm1)}\,|j\,m'\,(m\pm1)\rangle \rightarrow \hat{J}_\pm|j\,J\,m\rangle = \sqrt{J^2-m^2}\,|j\,J\,(m\pm1)\rangle \quad (2d)$$

$$\hat{j}_\mp|j\,m'\,m\rangle = \sqrt{j(j+1)-m'(m'\pm1)}\,|j\,(m'\pm1)\,m\rangle \rightarrow \hat{j}_\mp|j\,J\,m\rangle = 0 \qquad (2e)$$

We see that Eqs. (2a-e) are consistent with asymptotic forms of the left-column expressions for the $|j\,m'\,m\rangle$, with $m(m\pm1)\to m^2$ and $\sqrt{j(j+1)}\to J$. Note that $\hat{j}_\mp\,|j\,J\,m\rangle = 0$ always, consistent with $\sqrt{j(j+1)-m'(m'\pm1)} \to \sqrt{J^2-m'^2}$, which gives 0 for $m'=J$; we conclude that the length $|\mathbf{j}|$ and its projection along $z'$ (both $J$) are constant and therefore cannot be raised or lowered. For indicative derivations of some of Eqs. (2), see Appendix A.

## III. WAVEPACKETS

Quantum-state wavepackets are important in atomic, molecular, optical, and condensed matter physics, because they localize the underlying quantum system and elicit (semi)classical behavior from it. Thus, the formation and evolution of wavepackets has been extensively studied in various contexts, *e.g.*, position/momentum wavepackets of free particles [3], wavepackets of Rydberg states in atoms [22,23,24], coherent, squeezed, and intelligent states of light [1,25,26,27,28,29,30] and generally of angular momentum [31,32,33], or with a wider scope [11,34,35].

Here, the exponentials $e^{im\varphi}$ and $e^{i(j+\frac{1}{2})\chi}$ in $X_m^j(\varphi,\theta,\chi)$ allow the formation of spatial angular-momentum wavepackets $|X_{\Delta j,\Delta m}^{j,m}\rangle$, by summing the $|j'm'\rangle \equiv X_{m'}^j(\varphi,\theta,\chi)$ states over Gaussian distributions in $j'$ and $m'$, with widths of $\Delta j$ and $\Delta m$, respectively:

$$|X_{\Delta j,\Delta m}^{j,m}\rangle = \sum_{j',m'} e^{-\left(\frac{j'-j}{2\Delta j}\right)^2} e^{-\left(\frac{m'-m}{2\Delta m}\right)^2} X_{m'}^j(\varphi,\theta,\chi) \qquad (3)$$

The distributions of the conjugate-variable pairs $m,\varphi$ and $j,\chi$ are then described by the angular uncertainty relations [31], which for untruncated Gaussian distributions ($1 \ll \Delta j \ll j$, $1 \ll \Delta m \ll j$) take on the familiar form:

$$\Delta m\Delta\varphi = \hbar/2 \qquad (4a)$$

$$\Delta j\Delta\chi = \hbar/2 \qquad (4b)$$

Equations (4) apply exactly in the high-$j$ limit, where $\Delta\theta = 0$, as $\theta$ is constant and given by the VM expression $cos\theta = m/J$; for finite $j$, we take the derivative of this expression, and find that $\Delta\theta$ is:

$$\Delta\theta = \frac{\Delta m}{J sin\theta} \qquad (5)$$

Eq. (5) can be combined with Eq. (4a) to give an angular uncertainty relation for $\Delta\theta\Delta\varphi$:

$$(J\sin\theta)\,\Delta\theta\Delta\varphi = \hbar/2 \qquad (6)$$

Equations (4a) and (5) or (6) describe the angular size of the $j$-wavepacket, given by $\Delta\theta$ and $\Delta\varphi$. Note that $\theta$ and $\varphi$ are commuting variables, so that if there is no limit on the size of the angular momenta in the wavepacket ($J \to \infty$) then $\Delta\theta\Delta\varphi = 0$, and $\Delta\theta$ and $\Delta\varphi$ can each be arbitrarily small. However, if the average value the angular-momentum wavepacket is a finite value of $J$, then Eq. (6) gives the correlated uncertainty between $\Delta\theta$ and $\Delta\varphi$.

The spatial wavepacket of the associated particle position $|\psi_{\Delta j,\Delta m}^{j,m}\rangle$, for integer orbital-angular momentum $j$, is given by the sum of Eq. (3), but this time over spherical harmonics $|j'm'\rangle \equiv Y_{m'}^{j'}(\theta,\varphi)$:

$$|\psi_{\Delta j,\Delta m}^{j,m}\rangle = \sum_{j',m'} A e^{-\left(\frac{j'-j}{2\Delta j}\right)^2} e^{-\left(\frac{m'-m}{2\Delta m}\right)^2} Y_{m'}^{j'}(\theta,\varphi) \qquad (7)$$

where $A$ is normalization constant so that $\langle\psi_{\Delta j,\Delta m}^{j,m}|\psi_{\Delta j,\Delta m}^{j,m}\rangle = 1$, which, for $\Delta j, \Delta m \gg 1$, is given by $A = (2\pi \Delta j \, \Delta m)^{-1/2}$. In contrast to the $j$-wavepacket sum of Eq. (3), it is less clear that the particle-wavepacket sum of Eq. (7) will yield a localized wavepacket, because of the complicated $\theta$ dependence of the $Y_m^j(\theta,\varphi)$. However, the localization of the $j$-wavepacket leads

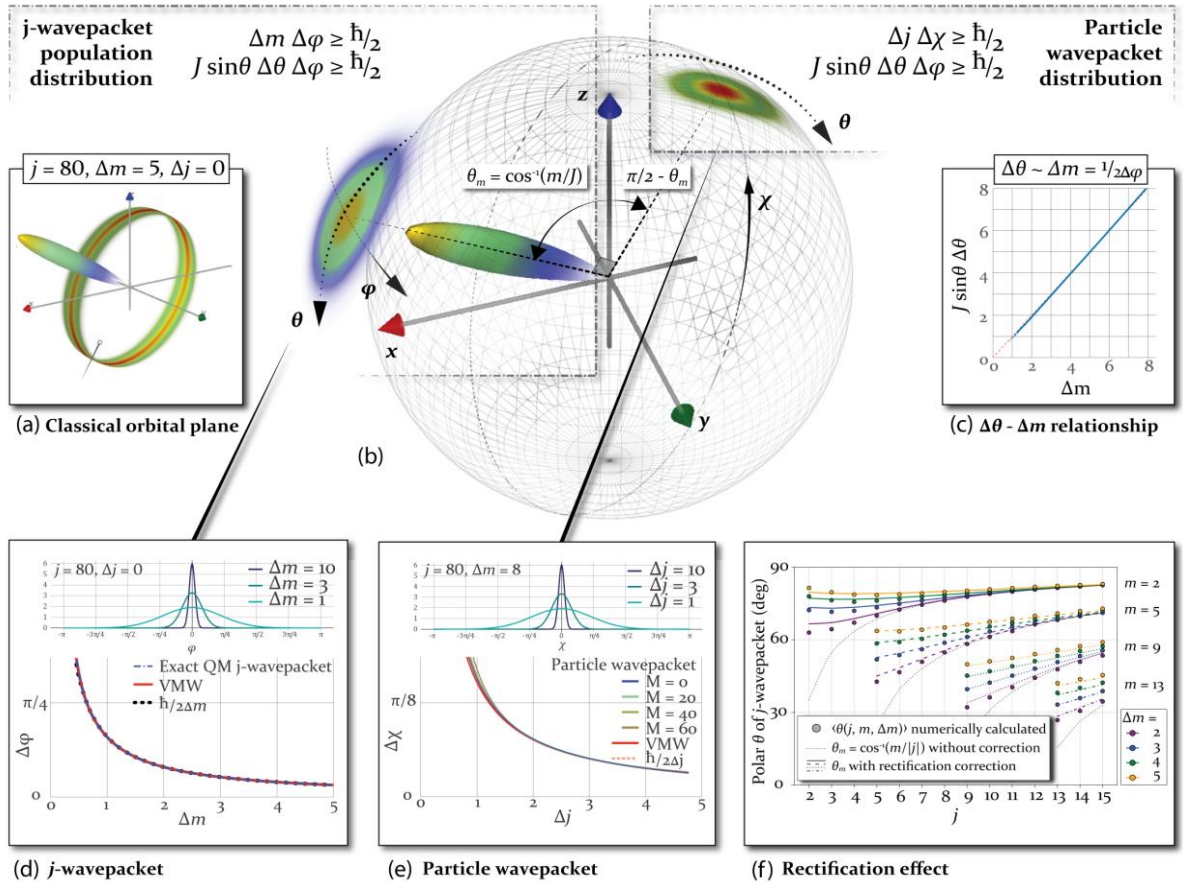

*Fig. 2: (a) Summing over $m'$ in Eqs. (3) and (7), for $j = 80$, $\Delta m = 5$, produces a $j$-wavepacket along polar orientation $\theta_m$ in the XZ plane, and the particle orbital plane emerges. (b) Summing over $j'$ in Eq. (7) localizes the particle in the XZ plane, at angle $\pi/2 - \theta_m$. Verification of the effective uncertainty relations for (c) $(J\sin\theta)\,\Delta\theta = \Delta m$ for $\Delta m \geq 1$, (d) $\Delta m\Delta\varphi \geq \hbar/2$ for $\Delta m \geq 1/2$, and (e) $\Delta j\Delta\chi \geq \hbar/2$ for $\Delta j \geq 2$. (f) Accounting for rectification of m-state distributions leads to corrected polar angles, $\tilde{\theta}_m$, which coincide well with actual values (see Appendix B).*

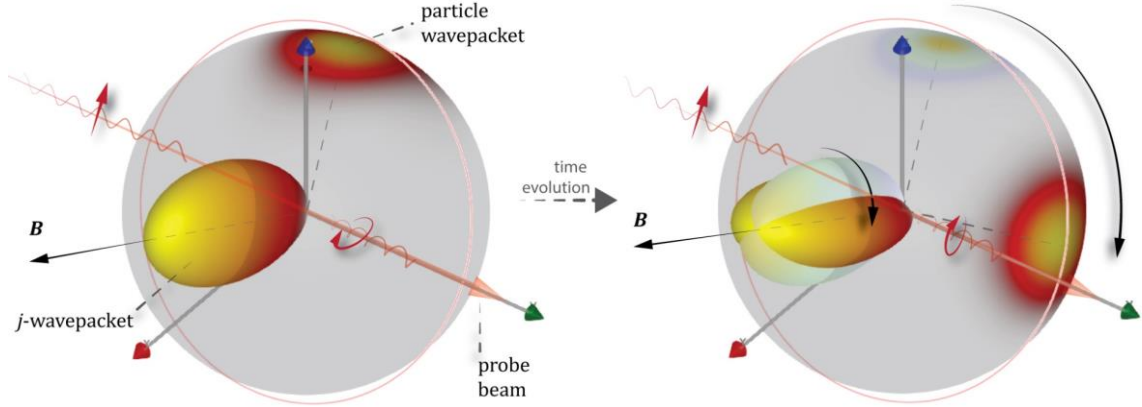

*Fig. 3: [Left to right] Time evolution of the j- and particle wavepackets as they precess about a magnetic field along the j-wavepacket axis, at $\theta_m = \cos^{-1}(m/|\mathbf{j}|)$. The j-wavepacket precession is mirrored by that of the particle wavepacket on the perpendicular, orbital χ-plane. Ellipsometric measurements on linearly polarized light traversing the j-wavepacket, would also indirectly measure the position of the particle wavepacket, without collapsing the particle wavefunction.*

us to expect that the angular position of the particle wavepacket will also be localized in a plane perpendicular to the average direction of the *j*-wavepacket, and its angular size will also be described by Eqs. (4-6). This expected localization indeed manifests and is shown in Fig. 2a, where the first sum over $m'$ in Eqs. (3) and (7) (for $\Delta m \gtrsim \sqrt{j}/2$) produces a *j*-wavepacket in the XZ plane and at polar angle $\theta_m$, and the particle is localized in the χ-plane, perpendicular to the *j*-wavepacket. Subsequently, the second summation over $j'$ in Eqs. (3) and (7) (for $\Delta j \gtrsim 2$) further localizes the particle in the XZ plane and at the polar angle $\pi/2 - \theta_m$ (see Fig. 2b). We determine $\Delta\theta$ and $\Delta\chi$ by fitting the particle wavepacket to Gaussians, and $\Delta\varphi$ is determined by calculating the expectation values of the operators $\langle\varphi^2\rangle$ and $\langle\varphi\rangle$ for the *j*-wavepacket density matrix [31]. In Fig. 2c-e, we show that Eqs. (4-6) are excellent approximations for $\Delta\varphi, \Delta\theta$, and $\Delta\chi$, and therefore show that the angular sizes of both wavepackets are predicted well by $m$, $\Delta m$, $j$, and $\Delta j$. These expressions hold well for $j \gtrsim 5$, but also give a good qualitative description even for lower $j$, as the average *j*-wavepacket remains perpendicular to the particle-wavepacket direction, and the Gaussian fits of the angular distributions for both the *j*- and particle-wavepackets remain excellent (Appendix B). We note that the *j*-wavepackets are illustrated by plotting the *population distribution* $Q(\theta, \varphi)$ of the maximum projection state along the direction defined by the angles $\theta, \varphi$ [36,37,38,39,40,41]. We also note an effect of the finite physical range of $m$ and $j$: when $m'$ and $j'$ in Eqs. (3) and (7) take non-physical values, we clamp the Gaussians to zero and the distributions become *rectified* [42]. This affects the statistical properties of the distributions and shifts the polar orientation angle of the wavepackets, which depends on the mean value of $m$. We use corrective formulas to account for this effect (see [42] and Appendix B), with good results, particularly for $\Delta m \gtrsim 3$, as shown in Fig. 2f.

Finally, we demonstrate that the rotational motions of the *j*- and particle-wavepackets are coupled: rotation of the *j*-wavepacket about its VM axis at $\theta_m = \cos^{-1}(m/|\mathbf{j}|)$ is matched by an equivalent rotation of the particle-wavepacket on the perpendicular χ-plane about the same axis. To induce rotation, we introduce a magnetic field **B** parallel to the *j*-wavepacket direction and assume degenerate orbital-angular-momentum *j* states (as is the case of H or Rydberg atoms, for a particular and large value of the principal quantum number, $n$). The magnetic field lifts the degeneracy of the $m$-states along **B**, and causes both wavepackets to precess around **B**. In Fig. 3, we plot calculations that show this uniform angular rotation. The energy splitting due to the magnetic field is linear in $m$ for low **B**, causing the wavepackets to rotate classically, *i.e.*, without spreading and reviving. Had the $j$-states been non-degenerate, the wavepackets would undergo collapses and revivals at regular, known intervals [33,35], however the coupling of the rotational motions of the two wavepackets would persist. The $j$-wavepacket rotation can be observed experimentally, through optical ellipsometric measurements on near-resonant, linearly polarized light propagating through the sample perpendicular to **B**, with its polarization normal to the plane defined by **B** and the propagation direction. Thus, measurements of the $j$-wavepacket distribution can yield the angular position of the particle indirectly and nondestructively (without wavefunction collapse).

## IV. CLEBSCH-GORDAN COEFFICIENTS

Wigner derived an expression for the average value of the square of neighboring Clebsch-Gordan (C-G) coefficients, valid at high-$j$, based on the traditional VM [43]:

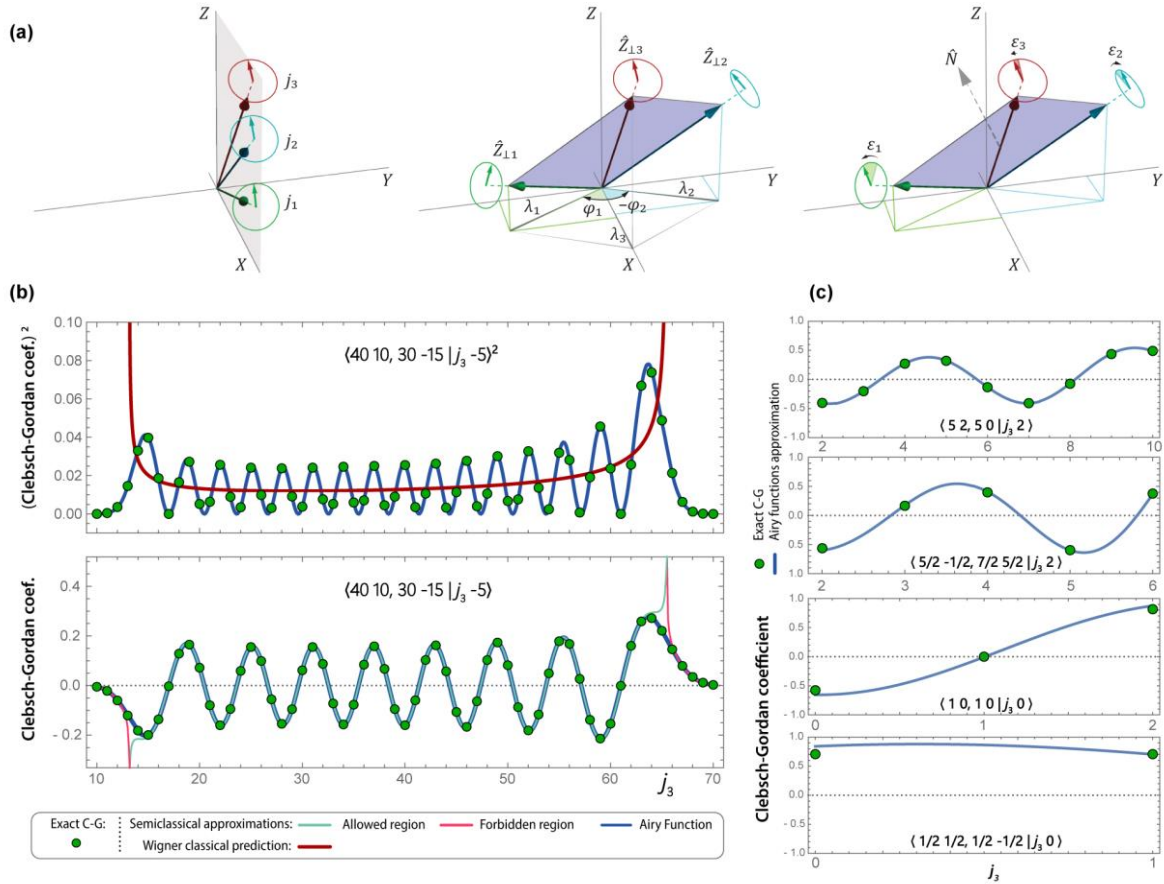

*Fig. 4: (a/left) VM wavefunctions zero-phase points: all vectors $\mathbf{j}_{1,2,3}$ lie in the XZ plane ($\varphi = 0$), and arrows about $\mathbf{j}_i$ point towards the Z-axis ($\chi = 0$). Rotation into coupling geometry: (a/middle) rotation about Z of $\mathbf{j}_1$ by $\varphi_1$ and $\mathbf{j}_2$ by $-\varphi_2$; (a/right) rotation of each $\mathbf{j}_i$ about the $j_i$-axis by angle $\varepsilon_i$. Comparisons of: (b/top) squares of C-G values to the classical expression of Eq. (8), for $j_1 = 40, m_1 = 10$, $j_2 = 30, m_2 = -15$, versus $j_3$; (b/bottom)C-G values to the VMW Eqs. (14, 15), for the allowed and forbidden regions. (c) VMW WKB solution (Eqs. (C9) and (C10) in Appendix C) for low-j, to C-G coefficients.*

$$\langle j_1 m_1, j_2 m_2 \mid j_3 m_3 \rangle^2_{avg.} = \frac{2(j_3+1)}{\pi \beta} \tag{8}$$

where $\beta$ is equal to four times the area of the triangle with sides $\lambda_1, \lambda_2, \lambda_3$, where the $\lambda_i = \sqrt{J_i^2 - m_i^2}$ are the projections of the $\boldsymbol{j_i}$ onto the XY plane (Fig. 4a), and $\beta$ is given by:

$$\beta = \sqrt{(\lambda_1 + \lambda_2 + \lambda_3)(-\lambda_1 + \lambda_2 + \lambda_3)(\lambda_1 - \lambda_2 + \lambda_3)(\lambda_1 + \lambda_2 - \lambda_3)} \tag{9}$$

We plot Eq. (8) against values of the squares of CG coefficients in Fig. 4b, showing typical behavior: $\langle j_1 m_1, j_2 m_2 \mid j_3 m_3 \rangle^2_{avg.}$ gives accurately the average value of neighboring squares of C-G coefficients, and the C-G coefficients oscillate about this average value. This behavior suggests a semiclassical expression for the C-G coefficients [18], of the form:

$$\langle j_1 m_1, j_2 m_2 \mid j_3 m_3 \rangle = F_{CVM} \times G_{VMphase} \tag{10}$$

where $F_{CVM} = \sqrt{2(j_3+1)/(\pi |\beta|)}$ is the classical coupling amplitude, given by the square root of Eq. (8) and $G_{VMphase}$ is determined from the tranformation properties of $j_m(\varphi, \theta, \chi)$ under rotation. The three $|j_1 m_1\rangle$, $|j_2 m_2\rangle$, $|j_3 m_3\rangle$ states are rotated into coupling position with the five rotation angles (Fig. 4a and Appendix C).

Multiplying all five rotation operators and the $\boldsymbol{j_1}$ - $\boldsymbol{j_2}$ exchange phase factor $e^{-i\Phi}$ yields:

$$\mathbf{R}_{j_1}(\varepsilon_1) \mathbf{R}_{j_2}(\varepsilon_2) \mathbf{R}_{j_3}(\varepsilon_3) \mathbf{R}_Z(\varphi_1) \mathbf{R}_Z(-\varphi_2) e^{-i\Phi} = e^{-i\Theta} \tag{11}$$

where $\Theta = \left(j_1 + \frac{1}{2}\right)\varepsilon_1 + \left(j_2 + \frac{1}{2}\right)\varepsilon_2 + \left(j_3 + \frac{1}{2}\right)\varepsilon_3 + m_1 \varphi_1 - m_2 \varphi_2 + \Phi$, with $\Phi = (j_3 - j_1 - j_2)\frac{\pi}{2}$, and $\mathbf{R}_N(\beta)$ is a rotation operator about direction $N$ by angle $\beta$ [10].

The classically allowed and forbidden regions are defined as $\Theta - \Phi$, $\beta \in \mathbb{R}$, $\beta > 0$ and $\Theta - \Phi$, $\beta \in \mathbb{I}$, respectively. Taking a normalized superposition of the two coupling geometries in the allowed region ($\pm\Theta$), and four in the forbidden region ($\Theta_{\pm\pm} = \pm Re[\Theta] \pm i(|Im[\Theta]| + \pi/4)$), gives, respectively (see Appendix C):

$$G_{VMphase}^{Allowed} = \frac{1}{\sqrt{2}}\left(e^{-i\Theta} + e^{i\Theta}\right) = \sqrt{2}\cos(\Theta) \tag{12}$$

$$G_{VMphase}^{Forbidden} = \frac{1}{2}\left(e^{i\Theta_{++}} + e^{i\Theta_{-+}} + e^{-i\Theta_{+-}} + e^{-i\Theta_{--}}\right) = 2\cos(Re[\Theta])e^{-|Im[\Theta]|-\pi/4} \tag{13}$$

Inserting Eqs. (12) or (13) into (10) yields the C-G coefficients [18, 44]:

**Allowed region:** $\langle j_1 m_1, j_2 m_2 \mid j_3 m_3 \rangle = 2\sqrt{\frac{(j_3+1)}{\pi\beta}}\cos(\Theta) \tag{14}$

**Forbidden region:** $\langle j_1 m_1, j_2 m_2 \mid j_3 m_3 \rangle = 2\sqrt{\frac{(j_3+1)}{\pi|\beta|}}\sqrt{2}\cos(Re[\Theta])\, e^{-|Im[\Theta]|-\pi/4} \tag{15}$

Schulten and Gordon showed that the intermediate turning-point regions are well-described with a WKB approximation solution involving Airy functions, which also gives the correct behavior in the allowed and forbidden regions [45]. We use similar expressions, Eqs. (C9) and (C10) in Appendix C, to compare to Eqs. (14) and (15) in Fig. 4b, and show excellent agreement even at low $j$ (Fig. 4c).

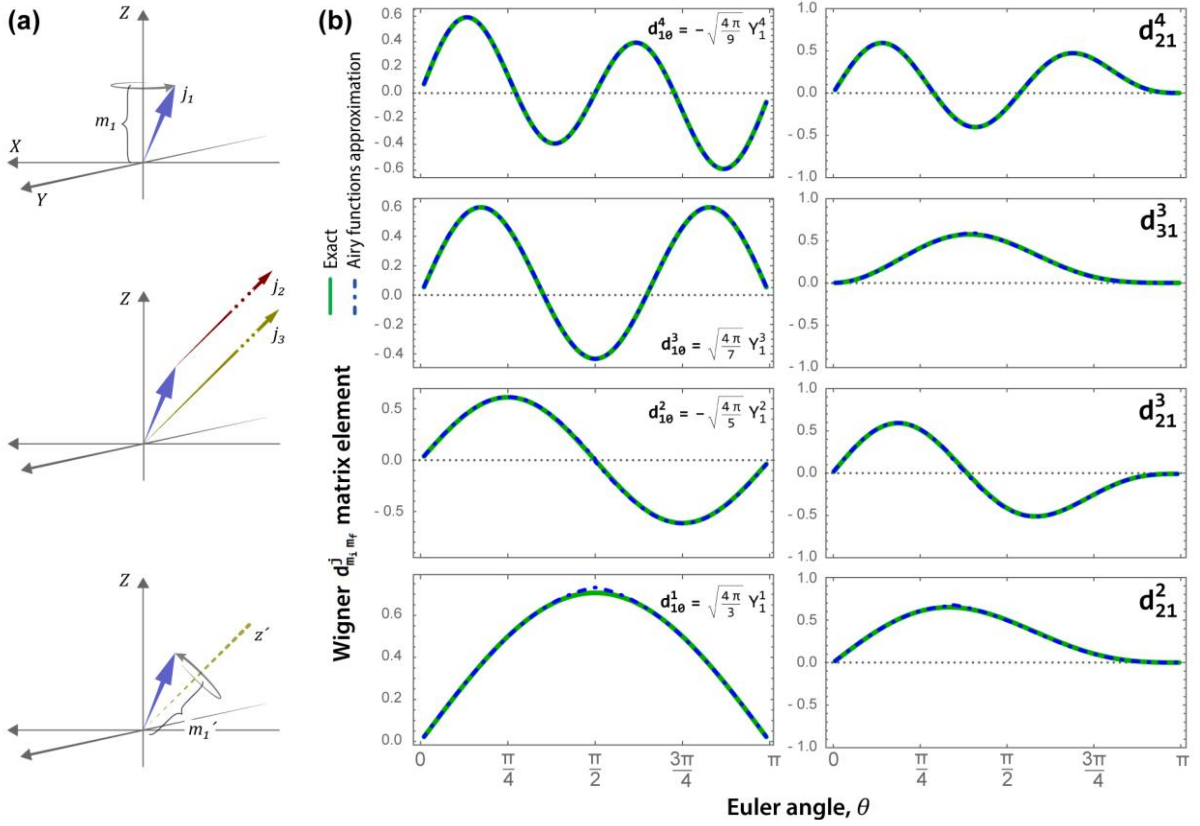

Fig. 5: (a) Coupling of finite $j_1$ to $j_2 \to \infty$, allows $d^{j_1}_{m'_1 m_1}$ to be expressed as the limit of a C-G coefficient. (b) The WKB expressions of Eqs. (D4) and (D5) of Appendix D are excellent approximations for the $d^j_{m'm}(\theta)$ elements, even for low $j$.

## V.  WIGNER ROTATION MATRIX ELEMENTS

We consider a state $|j_1 m_1\rangle$ with respect to the quantization axis Z. Then, any coupling to $Z$ is stopped (*e.g.*, any magnetic field along Z is turned off), and a large angular momentum $\boldsymbol{j_2}$ is introduced, which couples to $\boldsymbol{j_1}$ to give $\boldsymbol{j_3}$, with $j_3 = j_2 + m'_1$. In the limit of infinite $j_2$, as shown in Fig. 5, the C-G coefficient becomes equal to the Wigner rotation matrix element $d^{j_1}_{m'_1 m_1}(\theta)$ [46, 47], which describes the probability amplitude of the projection of $|j_1 m_1\rangle$ onto the state $|j_1 m'_1\rangle$, where the direction of $\boldsymbol{j_2}$ is the quantization axis:

$$d^{j_1}_{m'_1 m_1}(\theta) = (-1)^{j_1 - m'_1} \lim_{j_2 \to \infty} \langle j_1 m_1, j_2 m_2 \mid j_2 + m'_1, m_1 + m_2 \rangle \qquad (16)$$

where $\cos\theta = m_2/j_2$; note that $m_2 = j_2 \cos\theta$ also tends to infinity.

We now decouple $\boldsymbol{j_3}$ into $\boldsymbol{j_1}$ and $\boldsymbol{j_2}$, but along a quantization axis parallel to the infinite $\boldsymbol{j_2}$ (and $\boldsymbol{j_3}$), so that both are in maximal projection states $|j_2 j_2\rangle$ and $|j_3 j_3\rangle = |j_2 + m'_1, j_2 + m'_1\rangle$. Then, we see that decoupling occurs uniquely into the states $|j_1 m'_1\rangle$ and $|j_2 j_2\rangle$, because the decoupling coefficient is 1:

$$\lim_{j_2 \to \infty} \langle j_2 + m'_1, j_2 + m'_1 \mid j_2 j_2, j_1 m'_1 \rangle = 1 \qquad (17)$$

so that $\boldsymbol{j_1}$ is in the $|j_1 m'_1\rangle$ state along the $z'$ quantization axis (parallel to $\boldsymbol{j_2}$). Eq. (16) can be used to derive expressions for the $d^{j_1}_{m'_1 m_1}(\theta)$ (details are given in Appendix D). Again, WKB solutions involving Airy functions can be derived (see [48] and Appendix D), and excellent agreement between the VMW expressions and the exact $d^{j_1}_{m'_1 m_1}(\theta)$ is shown in Fig. 5 for low values of $j$.

## VI. CALCULATION OF g-FACTOR, $g = 2$

The magnetic moment of a particle is determined by the spatial distribution of its wavefunction. For example, an electron with mass $m$ and charge $e$, which is in the $|LM\rangle$ state, has a spatial wavefunction $\psi_{space} = Y^L_M(\theta, \varphi)$. The magnetic moment is then given by:

$$\hat{\mu}_Z = \left(\frac{e\hbar}{2m}\right)\hat{L}_Z \tag{18}$$

which yields a magnetic moment of $M\left(\frac{e\hbar}{2m}\right)$.

An elementary particle with mass $m$, charge $e$, and spin $S$, does not have a conventional spatial wavefunction analogous to the $Y^L_M(\theta, \varphi)$. However, we propose that the spatial distribution of the charge will transform as the angular momentum wavefunction $X^S_M(\varphi, \theta, \chi)$ (given by Eq. (1), with $J = S$), and therefore the spatial wavefunction of the particle will be given by $\psi_{space}(\varphi, \theta, \chi) = X^S_M(\varphi, \theta, \chi)$. The particle will have magnetic moment components $\hat{\mu}_Z$ and $\hat{\mu}_{z'}$, along the $Z$ and $z'$ axes, respectively, given by:

$$\hat{\mu}_Z = \left(\frac{e\hbar}{2m}\right)\hat{S}_Z \tag{19a}$$

$$\hat{\mu}_{z'} = \left(\frac{e\hbar}{2m}\right)\hat{S}_{z'} \tag{19b}$$

The $z'$ axis is distributed about $Z$ with cylindrical symmetry, so that only the parallel component of $\mu_{z'}$ contributes to the total magnetic moment along Z, $\hat{\mu}_{Z,\text{total}}$ :

$$\hat{\mu}_{Z,\text{total}} = \hat{\mu}_Z + \cos\theta_m \ \hat{\mu}_{z'} = \left(\frac{e\hbar}{2m}\right)\left(\hat{S}_Z + \cos\theta_m \hat{S}_{z'}\right) \tag{20}$$

The magnetic moment is then given by the eigenvalue of $\hat{\mu}_{Z,\text{total}}$ operating on $\psi_{space} = X^S_M(\varphi, \theta, \chi)$:

$$\hat{\mu}_{Z,\text{total}} \ \psi_{space} = (M + S\cos\theta_M)\left(\frac{e\hbar}{2m}\right)\psi_{space} = 2M\left(\frac{e\hbar}{2m}\right)\psi_{space} \tag{21}$$

where $\cos\theta_M = M/S$. The magnetic moment of an elementary particle is expressed generally as $gM\left(\frac{e\hbar}{2m}\right)$, so that Eq. (21) yields $g = 2$, independent of the value of $S$ (in contrast to the magnetic moment due to orbital angular momentum of an electron in the $|LM\rangle$ state, for which $g = 1$). The origin of $g = 2$ is clear here, with equal contributions to the total magnetic moment along Z from $\hat{\mu}_Z$ and $\hat{\mu}_{z'}$. The Dirac Equation and the Standard Model predict $g = 2$ (at the tree level) for spin ½ and 1, and several arguments are given that $g = 2$ for all elementary particles, independent of spin $S$ [49,50]. Interestingly, so does the VM wavefunction, without relativistic considerations [51,52].

## VII. M-STATE CORRELATIONS: $\langle j_3 m_3 \mid j_{1X} j_{2X} | j_3 m_3 \rangle$

The decomposition of $\boldsymbol{j}_3$ in the state $|j_3 m_3\rangle$, into $\boldsymbol{j}_1$ and $\boldsymbol{j}_2$, in terms of the product states

$|j_1 m_1\rangle|j_2 m_2\rangle$, is given by:

$$|j_3 m_3\rangle = \sum_{m_1} C_{m_2}^{m_1}|j_1 m_1\rangle|j_2 m_2\rangle \qquad (22)$$

where $C_{m_2}^{m_1} \equiv \langle j_1 m_1, j_2 m_2 \mid j_3 m_3 \rangle$ is an abbreviation for a C-G coefficient.

We show that the summation in Eq. (22) can be represented by a separate VM diagram, from which the expectation value of $\langle j_3 m_3 \mid j_{1X} j_{2X} | j_3 m_3 \rangle$ can be calculated geometrically. The azimuthal angle between the coupled $\boldsymbol{j}_1$ and $\boldsymbol{j}_2$ is $\varphi_{12}$ (Fig. 4a), given by (see Appendix E):

$$cos\varphi_{12} = \frac{j_3(j_3 + 1) - j_1(j_1 + 1) - j_2(j_2 + 1) - 2m_1(m_3 - m_1)}{2m_{1\perp}m_{2\perp}} \qquad (23)$$

and $m_{i\perp}$ is the projection of $\boldsymbol{j}_i$ in the XY plane. For particular values of projections $m_1$ and $m_2$, the product of $m_{1X}m_{2X}$ is determined geometrically using $m_{iX} = m_{i\perp}\cos(\varphi_i + \Phi)$, where $\Phi$ is the VM azimuthal delocalization angle. Summing over $m_1$, weighting each term by the coupling probability $\left[C_{m_2}^{m_1}\right]^2$, integrating over $\Phi$ and normalizing by $2\pi$, yields the VM value for $\langle j_3 m_3 \mid j_{1X} j_{2X} | j_3 m_3 \rangle$:

$$\langle j_3 m_3 \mid j_{1X} j_{2X} | j_3 m_3 \rangle = \frac{1}{2\pi} \int_0^{2\pi} \sum_{m_1} \left[C_{m_2}^{m_1}\right]^2 m_{1\perp} \cos(\varphi_1 + \Phi)\, m_{2\perp} \cos(\varphi_2 + \Phi)\, d\Phi \qquad (24)$$

Evaluating the integral in Eq. (24) gives:

$$\langle j_3 m_3 \mid j_{1X} j_{2X} | j_3 m_3 \rangle = \frac{1}{2} \sum_{m_1} \left[C_{m_2}^{m_1}\right]^2 m_{1\perp} m_{1X} \cos\varphi_{12} \qquad (25)$$

Finally, inserting Eq. (23) into (25) yields the result:

$$\langle j_3 m_3 \mid j_{1X} j_{2X} | j_3 m_3 \rangle = \frac{1}{4} \sum_{m_1} \left[C_{m_2}^{m_1}\right]^2 [j_3(j_3 + 1) - j_1(j_1 + 1) - j_2(j_2 + 1) - 2m_1 m_2] \qquad (26)$$

Although derived geometrically through the VM, Eq. (26) is the exact quantum mechanical result (see Appendix E).

## SUMMARY

In summary, we see that the extension of the Vector Model, the "Vector-Model" wavefunction, can be used to describe quantum-mechanical angular momentum exactly in the high-$j$ limit, with straightforward geometrical interpretations. Interestingly, the predictions remain excellent at low $j$ as well, suggesting that these geometrical interpretations are also useful at low $j$; very recent work has proposed exact low-$j$ wavefunctions, and investigated the usefulness of such wavefunctions [14]. The VMW wavepacket formalism simplifies the description of angular-momentum wavepackets and highlights the potential for experimental probing of particle-wavepacket rotation through continuous and non-destructive $\boldsymbol{j}$-rotation measurements.

## ACKNOWLEDGEMENTS

This work was partially supported by the Hellenic Foundation for Research and Innovation

(HFRI) and the General Secretariat for Research and Technology (GSRT), under grant agreement No. HFRI-FM17-3709 (project NUPOL). We thank Vasilis Niarchos, Dimitris Papazoglou, David Petrosyan, and Giorgos Vasilakis, and for reading the manuscript and making helpful suggestions.

**AUTHOR CONTRIBUTION STATEMENT**

T.P.R: Conceptualization, Methodology, Investigation, Supervision, Writing - Original Draft; M.E.K.: Investigation, Formal analysis, Writing - Review & Editing; G.E.K.: Investigation, Formal analysis, Validation, Visualization, Writing - Review & Editing.

**APPENDIX A, on Section II: VM WAVEFUNCTION**

**Indicative derivations of Equations (2)**

The operators $\hat{J}^2, \hat{J}_Z, \hat{J}_\pm$ in the space-fixed frame $XYZ$, and $\hat{J}^2, \hat{J}_{z'}, \hat{J}_\mp$ in the body-fixed frame $x'y'z'$, are given in terms of the Euler angles [10]. For the VM case of constant $\theta = \theta_m$, we give reduced operators, using units that give $\hbar = 1$:

$$\hat{J}^2 = -\frac{1}{\sin^2\theta_m}\left(\frac{\partial^2}{\partial\varphi^2} + \frac{\partial^2}{\partial\chi^2} - 2\cos\theta_m\frac{\partial^2}{\partial\varphi\partial\chi}\right) \tag{A1}$$

$$\hat{J}_Z = -i\frac{\partial}{\partial\varphi} \tag{A2}$$

$$\hat{J}_\pm = ie^{\pm i\varphi}\left[\cot\theta_m\frac{\partial}{\partial\varphi} - \frac{1}{\sin\theta_m}\frac{\partial}{\partial\chi}\right] \tag{A3}$$

$$\hat{J}_{z'} = -i\frac{\partial}{\partial\chi} \tag{A4}$$

$$\hat{J}_\mp = -ie^{\pm i\chi}\left[\cot\theta_m\frac{\partial}{\partial\chi} - \frac{1}{\sin\theta_m}\frac{\partial}{\partial\varphi}\right] \tag{A5}$$

**Derivation of Eq. (2a) of the main text:** operation of $\hat{J}^2$ on the $|j\,J\,m\rangle$ state:

$$\hat{J}^2|j\,J\,m\rangle = \frac{-1}{\sin^2\theta_m}\left(\frac{\partial^2}{\partial\varphi^2} + \frac{\partial^2}{\partial\chi^2} - 2\cos\theta_m\frac{\partial^2}{\partial\varphi\partial\chi}\right)e^{im\varphi}\delta(\theta - \theta_m)e^{iJ\chi}$$

$$= \left[\frac{1}{\sin^2\theta_m}\left(m^2 + J^2 - 2\left(\frac{m}{J}\right)mJ\right)\right]e^{im\varphi}\delta(\theta - \theta_m)e^{iJ\chi} = J^2\,|j\,J\,m\rangle \tag{A6}$$

where $\sin^2\theta_m = (J^2 - m^2)/J^2$. In addition, the state $|j\,m'\,J\rangle$ is also a solution (symmetric to $|j\,J\,m\rangle$ by the interchange of the Z and z' axes), as is the special case of the $|j\,J\,J\rangle$ state, for which $J$ and the Z and z' axes are all parallel (and $\theta_m = 0$).

**Derivation of Eq. (2d) of the main text:** operation of $\hat{J}_\pm$ on the $|j\,J\,m\rangle$ state

$$\hat{J}_\pm\,|j\,J\,m\rangle = ie^{\pm i\varphi}\left[\cot\theta_m\frac{\partial}{\partial\varphi} - \frac{1}{\sin\theta_m}\frac{\partial}{\partial\chi}\right]e^{im\varphi}\delta(\theta - \theta_m)e^{iJ\chi}$$

$$= \frac{-1}{\sin\theta_m}[m\cos\theta_m - J]e^{im\varphi}\delta(\theta - \theta_m)e^{i(m\pm1)\varphi} = \sqrt{J^2 - m^2}\,|j\,J\,(m\pm1)\rangle \tag{A7}$$

where $\sin\theta_m = \sqrt{J^2 - m^2}/J$, and $\delta(\theta - \theta_{m\pm1}) \to \delta(\theta - \theta_m)$ for m $\to \infty$.

**Derivation of Eq. (2e) of the main text:** operation of $\hat{J}_\mp$ on the $|j\,J\,m\rangle$ state:

$$\hat{J}_\mp\,|j\,J\,m\rangle = -ie^{\pm i\chi}\left[\cot\theta_m\frac{\partial}{\partial\chi} - \frac{1}{\sin\theta_m}\frac{\partial}{\partial\varphi}\right]e^{im\varphi}\delta(\theta - \theta_m)e^{iJ\chi}$$

$$= \frac{1}{\sin\theta_m}[J\cos\theta_m - m]e^{im\varphi}\delta(\theta - \theta_m)e^{i(J\pm1)\chi} = 0 \tag{A8}$$

where $\cos\theta_m = m/J$, so that $J\cos\theta_m - m = 0$.

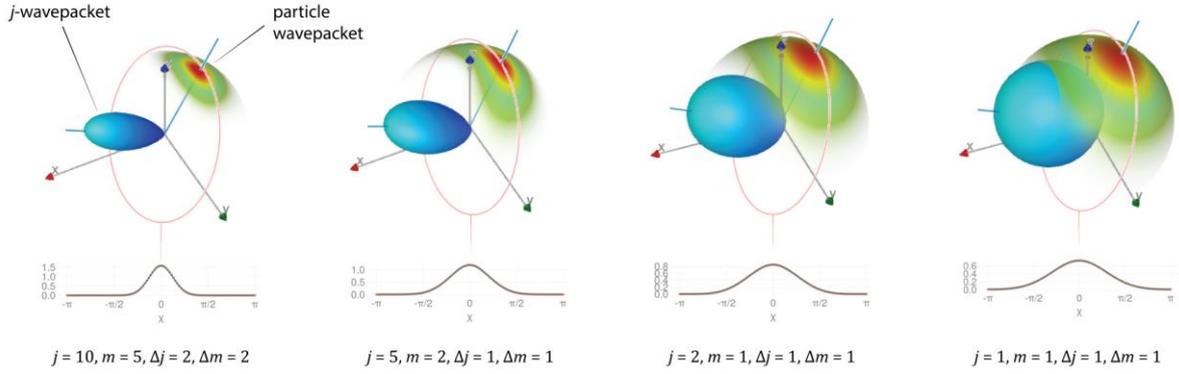

*Fig. A: The j-wavepacket, plotted with Eq. (3) via Eq. (B1) is perpendicular to the orbital plane of the particle wavepacket, plotted with Eq. (7) of the main text. Here plots for Δj = 1, and j values of (a) 10 (b) 5 (c) 2 (d) 1 are shown. Fitting to Gaussian distributions remains excellent, albeit the distributions are rectified for low j.*

## APPENDIX B, on Section III: WAVEPACKETS

### Angular momentum population distribution

The localization of the $j$- and particle wavepackets, and the emergence of classical characteristics such as the orbital plane of the particle motion, are discussed in the main article (see discussion regarding Fig. 2). The $j$-wavepackets are illustrated in Fig. 2 of the main text by plotting surfaces for which the distance from the axis origin is the *population distribution* magnitude, $Q(\theta, \varphi)$, of the maximum projection state along the direction defined by the angles $\theta, \varphi$ [36-40]:

$$Q(\theta, \varphi) = \sum_{k,q} \frac{2k+1}{4\pi} a_q^k \langle j\,j, k\,0 \mid j\,j \rangle\, C_q^{k*}(\theta, \varphi) \qquad (B1)$$

where the $a_q^k$ are the angular momentum polarization moments (*i.e.*, the wavepacket density-matrix spherical basis elements), the $C_q^k(\theta, \varphi)$ are reduced spherical harmonics [10], and $\langle j\,j, k\,0 \mid j\,j \rangle$ is a Clebsch-Gordan coefficient. For a thorough discussion of Eq. (B1), see [36-40]. These expressions hold very well for $j \gtrsim 5$, but also give a good qualitative description for even lower $j$, as the average $j$-wavepacket remains perpendicular to the particle orbital plane, and the Gaussian fits of the angular distributions for both the $j$- and particle-wavepackets remain excellent (Fig. A).

### Rectification of the Gaussian distributions

Both Fig. 2 and Fig. A show that the $j$-wavepacket is normal to the orbital plane of the particle wavepacket. However, in Fig. A, where low values of $j$ are used, the $j$-wavepacket is no longer oriented *exactly* at the Vector Model polar angle, $\theta_m = \cos^{-1}(m/|\boldsymbol{j}|)$. This deviation of $\theta_m$ is the result of the finite physical range of the $m$ quantum number for a given $j$: when the combination $m \pm$ few $\Delta m$ extends into non-physical values of $m$, the $m$-distribution becomes a *rectified* Gaussian distribution (*i.e.*, the Gaussian of Eq. (3) of the main text set to zero for $|m| > j$). This condition affects the statistical properties of the distribution (mean, variance etc.) and leads to a correction in $\theta_m$, which becomes significant when $m$ is close to the edge of its physical range and/or when $\Delta m$ is large. To estimate this correction, we first calculate the new mean value of the rectified $m$-distribution, using the formulas in [42]: for the lower and upper bounds of $m$, we define the transformed quantities: $a = \frac{-j-m}{\Delta m}$, $b = \frac{j-m}{\Delta m}$. We also define $g(x) = 1/\sqrt{2\pi}\, e^{-x^2/2}$ and remind the formula for the cumulative distribution function of the normal distribution, $\Phi(x) = \frac{1}{2}\left(1 + \text{erf}\left(\frac{x}{\sqrt{2}}\right)\right)$. Then, the mean and standard deviation of the rectified $m$-distribution are:

$$\bar{m} = m + \mu\,\Delta m \text{ and } \overline{\Delta m} = \sigma\,\Delta m \qquad (B2)$$

with $\mu = g(a) - g(b) + a\,\Phi(a) + b\,\Phi(-b)$, and

$$\sigma^2 = (\mu^2 + 1)\big(\Phi(b) - \Phi(a)\big) - (b - 2\mu)g(b) + (a - 2\mu)g(a) + (a - \mu)^2\Phi(a) + (b - \mu)^2\Phi(-b)$$

Using Eq. (B2) in the VM polar angle formula, $\bar{\theta}_m = \cos^{-1}(\bar{m}/|\boldsymbol{j}|)$, we calculate the corrected polar angle. In Fig. 2f of the main text we compare our estimates to the actual orientation angles, which were calculated numerically by maximizing the $j$-wavepacket distributions along the polar dimension, $\theta$, and for $\varphi = 0$, for each set of $(j, m, \Delta m)$ appearing in the Figure (circles). The rectification-corrected angles are plotted versus $j$ and are calculated for various $m$ (*line-style*-coded: solid, dashed, dotted, dot-dashed lines), and for various $\Delta m$ (*color-coded*: purple, blue, green, orange). We see that good agreement is achieved, particularly for values of $\Delta m \geq 3$, and $j \geq 5$.

## APPENDIX C, on Section IV: CLEBSCH-GORDAN COEFFICIENTS

### Coupling Geometry in detail

$G_{VMphase}$ is determined from the tranformation properties of $X_m^j(\varphi, \theta, \chi)$ under rotation. The wavefunctions of the three $|j_1 m_1\rangle$, $|j_2 m_2\rangle$, $|j_3 m_3\rangle$ states are in phase with respect to rotation about $Z$ (zero phase is defined by the X axis) and rotation about $\boldsymbol{j}_i$ (zero phase is defined by the Z axis). This means that we can draw the three $|j_1 m_1\rangle$, $|j_2 m_2\rangle$, $|j_3 m_3\rangle$ states, initially, as vectors in the XZ plane, with each $\boldsymbol{j}_i$ at angle $\theta_i$ to the Z axis. $G_{VMphase}$ is then determined from the following two transformations, which rotate the vectors $\boldsymbol{j}_1, \boldsymbol{j}_2, \boldsymbol{j}_3$ into the coupling geometry. Referring to Fig. 4 in the main text:

(1) The vectors $\boldsymbol{j}_1$ and $\boldsymbol{j}_2$ are rotated about the $Z$ axis to bring the three vectors into the coupling geometry determined by the vector addition, $\boldsymbol{j}_1 + \boldsymbol{j}_2 = \boldsymbol{j}_3$. Specifically, $\boldsymbol{j}_1$ is rotated by angle $\varphi_1$ about the $Z$ axis (described by the operator $\mathbf{R}_Z(\varphi_1) = e^{-im_1\varphi_1}$, where, generally, $\mathbf{R}_N(\beta)$ is a rotation operator about direction $N$ by angle $\beta$), and $\boldsymbol{j}_2$ is rotated in the opposite sense by angle $-\varphi_2$ (described by $\mathbf{R}_Z(-\varphi_2) = e^{im_2\varphi_2}$), see Fig. 4a, left and middle. The angle $\varphi_i$ is given by:

$$\varphi_i = \cos^{-1}\left(\frac{\lambda_i^2 + \lambda_3^2 - \lambda_j^2}{2\lambda_i\lambda_3}\right), \text{ for } i = 1, 2 \text{ and } i \neq j \qquad (C1)$$

(2) A phase-matching condition is achieved by rotating $\boldsymbol{j}_i$ vectors by angle $\varepsilon_i$ about the $\boldsymbol{j}_i$ axis, so that the zero phase points are rotated from $\widehat{\boldsymbol{Z}}_{\perp i}$ to the $\widehat{\boldsymbol{N}}$ axis (see Fig. 4a, right), perpendicular to the coupling plane. These three rotations are described by the operator $\mathbf{R}_{\boldsymbol{j}_i}(\varepsilon_i) = e^{-i\left(j_i + \frac{1}{2}\right)\varepsilon_i}$, and the angles $\varepsilon_i$ (for $i = 1, 2$) and $\varepsilon_3$ are given by:

$$\varepsilon_i = S_i \cos^{-1}\left(\frac{2}{\alpha}\lambda_3 j_i' \sin\varphi_i\right), \qquad \text{for } i = 1, 2 \tag{C2}$$

$$\varepsilon_3 = S_3 \cos^{-1}\left(\frac{2}{\alpha}\lambda_1 j_3' \sin\varphi_1\right) \tag{C3}$$

where $\alpha$ is equal to four times the area of the triangle with sides $j_1', j_2', j_3'$, given by:

$$\alpha = \sqrt{(j_1' + j_2' + j_3')(-j_1' + j_2' + j_3')(j_1' - j_2' + j_3')(j_1' + j_2' - j_3')} \tag{C4}$$

$S_i$ is the sense of rotation of angle $\varepsilon_i$ ($S_i$ takes values of $+1$ or $-1$), given by:

$$S_i = \frac{\left(\widehat{\boldsymbol{N}}_i \times \widehat{\boldsymbol{Z}}_{\perp i}\right) \cdot \boldsymbol{j}_i}{\left|\left(\widehat{\boldsymbol{N}}_i \times \widehat{\boldsymbol{Z}}_{\perp i}\right) \cdot \boldsymbol{j}_i\right|} \tag{C5}$$

where $\boldsymbol{N}_{1,2} = -\boldsymbol{N}_3 = \boldsymbol{j}_1 \times \boldsymbol{j}_2$ is a vector perpendicular to the coupling plane, and

$$\boldsymbol{Z}_{\perp i} = \widehat{\boldsymbol{Z}} - (\widehat{\boldsymbol{Z}} \cdot \boldsymbol{j}_i)\boldsymbol{j}_i \tag{C6}$$

Multiplying all five rotation operators yields:

$$\mathbf{R}_{j_1}(\varepsilon_1)\mathbf{R}_{j_2}(\varepsilon_2)\mathbf{R}_{j_3}(\varepsilon_3)\mathbf{R}_Z(\varphi_1)\mathbf{R}_Z(-\varphi_2) = e^{-i\Omega} \tag{C7}$$

where $\Omega = \left(j_1 + \frac{1}{2}\right)\varepsilon_1 + \left(j_2 + \frac{1}{2}\right)\varepsilon_2 + \left(j_3 + \frac{1}{2}\right)\varepsilon_3 + m_1\varphi_1 - m_2\varphi_2$.

Introducing also the $\boldsymbol{j}_1$ - $\boldsymbol{j}_2$ exchange phase factor $e^{-i\Phi}$, where $\Phi = (j_3 - j_1 - j_2)\pi/2$, needed to take into account possible sign changes from the exchange of $\boldsymbol{j}_1$ and $\boldsymbol{j}_2$, we get Eq. (11) of the main text:

$$\mathbf{R}_{j_1}(\varepsilon_1)\mathbf{R}_{j_2}(\varepsilon_2)\mathbf{R}_{j_3}(\varepsilon_3)\mathbf{R}_Z(\varphi_1)\mathbf{R}_Z(-\varphi_2)e^{-i\Phi} = e^{-i\Omega}e^{-i\Phi} = e^{-i\theta} \tag{C8}$$

where $\theta = \Omega + \Phi = \left(j_1 + \frac{1}{2}\right)\varepsilon_1 + \left(j_2 + \frac{1}{2}\right)\varepsilon_2 + \left(j_3 + \frac{1}{2}\right)\varepsilon_3 + m_1\varphi_1 - m_2\varphi_2 + (j_3 - j_1 - j_2)\frac{\pi}{2}$.

**WKB Solutions in detail**

*Allowed region* ($\Omega, \beta \in \mathbb{R}, \ \beta > 0$):

In the allowed region, there are two coupling geometries that are mirror images of each other, reflected through the $Z - \boldsymbol{j}_3$ plane, for which all the transformation angles have opposite signs, so that two geometries are defined by $+\Omega$ and $-\Omega$. $G_{VMphase}$ is then given by a normalized linear superposition of the transformation to these two coupling geometries. However, note that under exchange of $j_1$ and $j_2$, $\Omega$ changes sign, whereas the C-G coefficient changes by the phase $(-1)^{j_1+j_2-j_3}$ (as is well known, *e.g.*, [10] and [53]), and can be shown using $j_m(\varphi, \theta, \chi)$ and $\hat{J}_-$. To keep $G_{VMphase}$ and the C–G coefficients real, we share this phase between the two terms, by multiplying each term by $e^{-i\Phi}$ and $e^{i\Phi}$, respectively (where $\Phi = (j_3 - j_1 - j_2)\pi/2$ is an integer multiple of $\pi/2$), so that $G_{VMphase}$ and the semiclassical C-G coefficients have the correct symmetry under the exchange of $j_1$ and $j_2$ [10], and satisfy the standard C-G phase conventions (e.g. that the C-G coefficients are positive for $j_1 + j_2 - j_3 = 0$). Inserting Eq. (12) into Eq. (10) of the main text yields the semiclassical expression for C-G coefficients in the allowed region [18, 45], given by Eq. (14). This expression is equivalent to that given in the literature found heuristically [18] or derived by taking the asymptotic limit of C-G coefficients [44], although it appears slightly different due to the use of complementary angles.

***Forbidden region ($\Omega, \beta \in \mathbb{I}$):***

In the forbidden region, all the angles become complex and $\Theta$ has both real and imaginary parts: $Re[\Theta]$ and $Im[\Theta]$. In addition, $\beta$ is imaginary, so $\beta = \pm i|\beta| = e^{\pm i\pi/2}|\beta|$, and $1/\sqrt{\beta} = e^{\pm\frac{i\pi}{4}}/\sqrt{|\beta|}$. There are now four symmetric coupling geometries (the four combinations of the signs of the real and imaginary parts), described by: $\Theta_{\pm\pm} = \pm Re[\Theta] \pm i(|Im[\Theta]| + \pi/4)$. Taking the superposition of all four transformations gives Eq. (13). Inserting Eq. (13) into Eq. (10), and using $F_{CVM} = \sqrt{2(j_3+1)/\pi|\beta|}$, yields Eq. (15), which is equivalent to the asymptotic limit of C-G coefficients [44], however the sign is given geometrically by the term $cos(Re[\Theta])$, as opposed to non-geometrical and conditional expressions [44].

***Turning point region ($\beta^2 \simeq 0$):***

At the intermediate turning point regions, the solutions of Eqs. (14) and (15) diverge. Schulten and Gordon showed that these regions are well-described with a uniform WKB approximation solution involving Airy functions, which also gives the correct behavior in the allowed and forbidden regions [45]. We use similar expressions and compare them to Eqs. (14) and (15), showing excellent agreement even at low $j$, as seen in Fig. 4b,c. Our adapted expressions are given below. We define the phase factors:

$\Omega_o = \frac{\pi}{2} \times \left\{ J_1 S_1 + J_2 S_2 + J_3 S_3 + m_1[1 - \text{sgn}(\pi/2 - \text{Re}(\varphi_1))] - m_2[1 - \text{sgn}(\pi/2 - \text{Re}(\varphi_2))] \right\}$ and $\Delta_o = (J_1 + J_2 + J_3)\frac{\pi}{2}$.

Then the WKB approximation solution for the classically allowed and forbidden regions is:

***Allowed region ($\Omega, \beta \in \mathbb{R}, \ \beta > 0$):***

$$\langle j_1 m_1, j_2 m_2 \mid j_3 m_3 \rangle = (-1)^{j_1 - j_2}\sqrt{2j_3 + 1}\,\frac{Z^{1/4}}{\sqrt{\beta/2}} \times \begin{cases} c_o Ai(-Z) - s_o Bi(-Z), & \text{if } Re(\Omega - \Omega_o) < 0 \\ c_o Bi(-Z) - s_o Ai(-Z), & \text{if } Re(\Omega - \Omega_o) > 0 \end{cases} \quad \text{(C9)}$$

***Forbidden region ($\Omega, \beta \in \mathbb{I}$):***

$$\langle j_1 m_1, j_2 m_2 \mid j_3 m_3 \rangle = (-1)^{j_1 - j_2}\sqrt{2j_3 + 1}\,\frac{Z^{1/4}}{\sqrt{|\beta|/2}} \times \begin{cases} c_o Ai(Z) - s_o Bi(Z), & \text{if } Im(\Omega - \Omega_o) > 0 \\ c_o Bi(Z) - s_o Bi(Z), & \text{if } Im(\Omega - \Omega_o) < 0 \end{cases} \quad \text{(C10)}$$

where $Z = (3|\Omega - \Omega_o|/2)^{2/3}$, and $c_o = \cos(\Omega_o + \Delta_o)$ and $s_o = \sin(\Omega_o + \Delta_o)$. We note that for any physically allowed set of quantum numbers, $j_1, m_1, j_2, m_2, j_3,$ and $m_3$, only one of the branching parameters $c_o$ or $s_o$ is non-zero and equal to $\pm 1$. We also note that these formulas are only valid for integer $j_3$ (but $j_1, j_2$ either integers or half-integers). For half-integer $j_3$ the Clebsch-Gordan symmetry relations can be used:

$$\langle j_1 m_1, j_2 m_2 \mid j_3 m_3 \rangle = (-1)^{j_1 - m_1}\frac{\sqrt{2j_3 + 1}}{\sqrt{2j_2 + 1}}\, \langle j_1 m_1, j_3 - m_3 \mid j_2 - m_2 \rangle \quad \text{(C11)}$$

$$= (-1)^{j_2 - m_2}\frac{\sqrt{2j_3 + 1}}{\sqrt{2j_1 + 1}}\, \langle j_3 - m_3, j_2 m_2 \mid j_1 - m_1 \rangle \quad \text{(C12)}$$

## APPENDIX D, on Section V: WIGNER ROTATION MATRIX ELEMENTS

### Uniform WKB Solution

In Section V of the main text, using the VM, we show that the amplitude of the projection of $|j_1 m_1\rangle$ onto $z'$ (at angle $\theta$ to $Z$ and in the $X$-$Z$ plane) to give $|j_1 m'_1\rangle$, given by $d^{j_1}_{m'_1 m_1}(\theta)$, is consistent with the limiting case of a C-G coefficient, as expressed in Eq. (16) of the main paper. We now insert the C-G coefficient expressions from Section IV into Eq. (16) and evaluate the limit to produce asymptotic expressions for the $d^j_{m'm}(\theta)$. We define he quantity:

$$R \equiv R(m, m') \equiv J^2 \sin^2\theta - m^2 - m'^2 + 2mm' cos\theta \tag{D1}$$

in terms of which we separate the classically allowed region, for which $R > 0$, from the forbidden, for which $R < 0$ [48]. The asymptotic solutions for the two regions are found to be:

*Allowed region* $(R > 0)$:

$$d^j_{m'm}(\theta) = \frac{\sqrt{\frac{2}{\pi}}}{R^{\frac{1}{4}}} \cos(\theta) \tag{D2}$$

*Forbidden region* $(R < 0)$:

$$d^j_{m'm}(\theta) = \frac{\sqrt{\frac{2}{\pi}}}{|R|^{\frac{1}{4}}} \sqrt{2} \cos(Re[\theta]) \ e^{-|Im[\theta]| - \frac{\pi}{4}} \tag{D3}$$

where

$$\theta = -m' \cos^{-1}\left[\frac{m - m' cos\theta}{sin\theta \sqrt{J^2 - m'^2}}\right] + J \cos^{-1}\left[\frac{mm' - J^2 cos\theta}{\sqrt{J^2 - m^2}\sqrt{J^2 - m'^2}}\right] + m \cos^{-1}\left[\frac{m cos\theta - m'}{sin\theta \sqrt{J^2 - m^2}}\right] - \frac{\pi}{4} \tag{D4}$$

and $J = j + \frac{1}{2}$. At the boundary of the classically allowed domain $(R = 0)$, the simple expressions of Eqs. (D2) and (D3) diverge, and, as in the case of the Clebsch-Gordan coefficients, a uniform WKB solution involving Airy functions can be found that describes both regions and joins them smoothly at the turning points [48]. This solution then provides excellent approximations for the $d^j_{m'm}(\theta)$ elements, even for low values of $j$, as shown in Fig. 5. The WKB solution for the $d^j_{m'm}(\theta)$ is given below for $0 < \theta < \pi/2$, $0 < m$, and $|m'| < m$:

$$d^j_{m'm}(\theta) = \left(-\frac{4Z}{R}\right)^{1/4} \text{Ai}(Z), \qquad \text{for m}' < m\cos\theta \tag{D4}$$

$$d^j_{m'm}(\theta) = (-1)^{j-m}\left(-\frac{4Z}{R}\right)^{1/4} \text{Ai}(Z), \qquad \text{for m}' \geq m\cos\theta \tag{D5}$$

where Z is given in the *allowed region* by:

$$Z = -\left[\frac{3}{2}\left(J\pi - (\theta + \pi/4)\right)\right]^{2/3}, \qquad \text{for } m' < m\cos\theta \tag{D6}$$

$$Z = -\left[\frac{3}{2}(\theta + \pi/4 - m\pi)\right]^{2/3}, \qquad \text{for } m' \geq m\cos\theta \tag{D7}$$

and in the **forbidden region** by:

$$Z = \left[\frac{3}{2}\text{Im}(-\theta)\right]^{2/3} \tag{D8}$$

with

$$\text{Im}(-\theta) = \mp m' \cosh^{-1}\left[\frac{m-m'\cos\theta}{\sin\theta\sqrt{J^2-m'^2}}\right] - J\cosh^{-1}\left[\frac{|J^2\cos\theta-mm'|}{\sqrt{J^2-m^2}\sqrt{J^2-m'^2}}\right] + m\cosh^{-1}\left[\frac{|m\cos\theta-m'|}{\sin\theta\sqrt{J^2-m^2}}\right] \tag{D9}$$

where, in the first term, the "$-$" and "$+$" signs are for $m' <$ and $\geq m\cos\theta$, respectively.

The remaining parameter regions are captured by the symmetry relations:

$$d^{j}_{m'm}(\theta) = d^{j}_{-m'-m}(\theta) = (-1)^{m'-m}d^{j}_{mm'}(\theta) = (-1)^{m'-m}d^{j}_{m'm}(-\theta) = (-1)^{j-m}d^{j}_{-m'm}(\pi-\theta) \tag{D10}$$

## APPENDIX E, on Section VII: M-STATE CORRELATIONS

### Details on the derivation of Eq. (23)

The decomposition of $\boldsymbol{j_3}$ in the state $|j_3m_3\rangle$ into $\boldsymbol{j_1}$ and $\boldsymbol{j_2}$ in terms of the product states $|j_1m_1\rangle|j_2m_2\rangle$, as stated in Eq. (22) of the main text. Usually, a single VM diagram, which is merely the classical vector coupling diagram that precesses about the Z axis, is used to describe qualitatively all the terms in Eq. (22) [10,4]. Here, we show that a separate VM pictorial representation for *each* term in Eq. (22), allows the correct calculation of $\langle j_3m_3 \mid j_{1X}j_{2X}|j_3m_3\rangle$, which further extends the usefulness of the VM (see Fig. B).

For each term in Eq. (22), $\boldsymbol{j_1}$ and $\boldsymbol{j_2}$ have a constant relative azimuthal angle between them, $\varphi_{12}(m_1)$. All three vectors, $\boldsymbol{j_1}$, $\boldsymbol{j_2}$, and $\boldsymbol{j_3}$, are fixed relative to each other, but are delocalized about

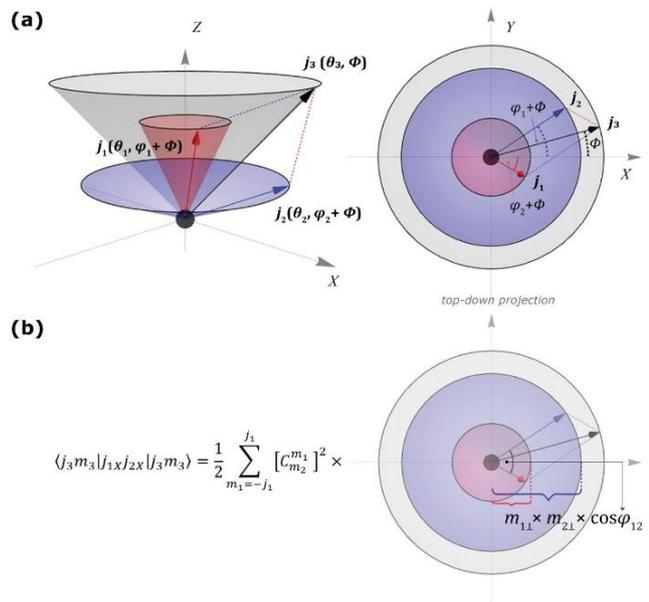

*Fig. B: (a) The VM pictorial representation of each term of the decomposition of the state $|j_3m_3\rangle$ (Eq. 21). (b) Pictorial calculation of $\langle j_3m_3 \mid j_{1X}j_{2X}|j_3m_3\rangle$, using Eq. (25).*

Z together. For each VM figure, the relative projections for $\boldsymbol{j_1}$ and $\boldsymbol{j_2}$, on all three $X$, $Y$, and $Z$ axes, can be determined, so that the correlations of the projections, $\langle j_3m_3 \mid j_{1X}j_{2X}|j_3m_3\rangle$ or $\langle j_3m_3 \mid j_{1Y}j_{2Y}|j_3m_3\rangle$, can be calculated pictorially, by integrating over the delocalization angle. The magnitude of $\boldsymbol{j_3}$ is given by:

$$|\boldsymbol{j_3}|^2 = |\boldsymbol{j_1}+\boldsymbol{j_2}|^2 = |\boldsymbol{j_1}|^2 + |\boldsymbol{j_2}|^2 + 2|\boldsymbol{j_1}||\boldsymbol{j_2}|\cos\theta_{12} \tag{E1}$$

where $\cos\theta_{12}$ is the angle between $\boldsymbol{j_1}$ and $\boldsymbol{j_2}$, which is expressed in terms of $\theta_1$, $\theta_2$, and $\varphi_{12}$ as:

$$\cos\theta_{12} = \cos\theta_1\cos\theta_2 + \sin\theta_1\sin\theta_2\cos\varphi_{12} \tag{E2}$$

with $cos\theta_i = m_i/|\boldsymbol{j}_i|$ and $sin\theta_i = m_{i\perp}/|\boldsymbol{j}_i|$, where $m_{i\perp}$ is the projection of $\boldsymbol{j}_i$ on the XY plane.

Inserting Eq. (E2) into (E1) and solving for $cos\varphi_{12}$ yields Eq. (23) of the main text.

The remaining steps of the calculation are discussed in the main text, leading to Eq. (26), which, although derived pictorially, is the exact quantum mechanical result. The latter is also calculated straightforwardly, by using Eq. (22) and

$$j_{1X}j_{2X} = (j_3^2 - j_1^2 - j_2^2 - 2j_{1Z}j_{2Z} + j_{1+}j_{2+} + j_{1-}j_{2-})/4$$

Similar procedures give the same result for $\langle j_3 m_3 \mid j_{1Y}j_{2Y}|j_3 m_3\rangle$.